\documentstyle[11pt,fleqn,epsf]{article}

\newcommand{\beq}{\begin{equation}}
\newcommand{\eeq}{\end{equation}}
\newcommand{\ba}{\begin{eqnarray}}
\newcommand{\ea}{\end{eqnarray}}
\renewcommand{\a}{\alpha}

\newcommand{\m}{\mu}
\newcommand{\n}{\nu}

\newcommand{\text}{\textstyle}
\renewcommand{\sc}{\scriptstyle}
\begin{document}

\begin{titlepage}
\hfill THES-TP 98/2
\begin{center}

\LARGE

{\bf{Finite-Size Effects and Operator Product Expansions in a
CFT for $d>2$}} \vspace{5cm} \normalsize

{\bf{Anastasios C. Petkou }}\footnote{e-mail:
apetk@skiathos.physics.auth.gr} and {\bf {Nicholas D.  Vlachos }}
\footnote{e-mail: vlachos@physics.auth.gr}\\

Department of Physics \\Aristotle University of Thessaloniki\\Thessaloniki
54006, Greece
\end{center}
\vspace{2cm} \centerline{{\bf{Abstract}}}

\vspace{0.7cm}

The large momentum expansion for the inverse propagator of the auxiliary
field $\lambda(x)$ in the conformally invariant $O(N)$ vector model is
calculated to leading order in $1/N$, in a strip-like geometry with one
finite dimension of length $L$ for $2<d<4$. Its leading terms are
identified as contributions from $\lambda(x)$ itself and the energy
momentum tensor, in agreement with a previous calculation based on conformal
operator product expansions. It is found that a  non-trivial
cancellation takes place by virtue of the gap equation. The leading
coefficient of the energy momentum tensor contribution is shown to be
related to the free energy density.

\end{titlepage}

The effects of finite geometry in systems near  second order phase
transitions points are of great importance for statistical mechanics and
quantum field theory. These effects are largely explained by the theory of
finite size scaling \cite{Zinn-Justin}. On the other hand, from a purely
field theoretical point of view, second order phase transitions are
connected to conformal field theories (CFTs). In $d=2$ spacetime
dimensions, there is an almost complete understanding of CFTs in terms of
operator product expansions (OPEs) \cite{Ginsparg}. Recently, there has
been some progress towards understanding CFTs for $d>2$ in terms of OPEs
plus some additional algebraic conditions such as the cancellation of
``shadow singularities'' \cite{Ruhl, Tassos1, Tassos2}. The CFT approach
to finite size scaling at criticality is based on the observation that the
OPE, being a short distance property of the critical theory, is
insensitive to finite size effects. This means, for example, that the
finite size corrections of two-point functions are directly related to
specific terms in the OPE. Such an approach has been very
successful when applied in $d=2$ \cite{Cardy1}.

In the present work, the conformally invariant $O(N)$ vector model is
investigated in $2<d<4$, in order to demonstrate that OPE techniques can
be useful for studying finite size effects of critical systems in $d>2$ as
well. The geometry is taken to be strip-like with one finite dimension of
length $L$ and periodic boundary conditions. In this case, for $2<d\leq 3$
the critical gap equation has only one ``massive'' solution, while,
for $3<d<4$ the gap equation has both a ``massive'' and a ``massless''
solution. We study the large momentum expansion for the inverse
propagator of the auxiliary field $\lambda (x)$, which is equivalent to an
OPE. The leading scalar and tensor contributions to this OPE for the
``massive'' and ``massless'' solutions of the critical gap equation above
are identified. In both cases the results are in agreement with what would
be expected from abstract CFT in $d>2$. Most importantly, for the
``massive'' solution of the gap equation, a non-trivial cancellation is
found to take place inside the inverse propagator of $\lambda (x)$. This
is similar to the ``shadow singularities'' cancellation found in
\cite{Tassos1, Tassos2}. Finally, it is shown that the leading coefficient
of the energy momentum tensor $T_{\mu\nu}(x)$ contribution is related to
the free energy density of the model and also to $C_{T}$, the latter being
the overall universal scale in the two-point function of $T_{\mu\nu}(x)$.

The Euclidean partition function of the $O(N)$ vector model is given by
\begin{equation}
Z=\int ({\cal {D}}\phi ^{\alpha })({\cal {D}}\sigma )\,\exp \left[
-\frac{1}{ 2}\int {\rm d}^{d}x\bigl[\phi ^{\alpha }(x)\bigl(-\partial
_{x}^{2}+\sigma (x)\bigl)\phi ^{\alpha }(x)\bigl]+\frac{1}{2f}\int {\rm
d}^{d}x\,\sigma (x) \right] \,,  \label{eq1}
\end{equation}
where $\phi
^{\alpha }(x)$, $\alpha =1,..,N$ is the basic $O(N)$-vector, $
O(d)$-scalar field, $\sigma (x)$ is the $O(d)$-scalar auxiliary field and
$f$ is the coupling. Integrating out $\phi ^{a}(x)$, we obtain
\begin{equation}
Z=\int ({\cal {D}}\sigma )\,\exp \left[ -\frac{N}{2}S_{eff}(\sigma ,g)\right]
,\,\,\,S_{eff}(\sigma ,g)=\mbox{Tr}[\ln (-\partial ^{2}+\sigma )]-\frac{1}{g}
\int {\rm d}^{d}x\,\sigma (x)\,,  \label{eq2}
\end{equation}
for the rescaled coupling $g=Nf$. Setting $\sigma
(x)=m^{2}+(i/\sqrt{N})\lambda (x)$, where $m^{2}$ is the stationary value
of the functional integral (\ref{eq2}) given by the gap equation
\begin{equation}
\int \frac{{\rm d}^{d}p}{(2\pi
)^{d}}\frac{1}{p^{2}+m^{2}}=\frac{1}{g}\,, \label{eq3}
\end{equation}
and expanding (\ref{eq2}) in powers of $\lambda(x)$, we
get the usual large $N$ expansion. The momentum space inverse propagator of
$\lambda (x)$ is
\begin{equation} \Pi ^{\text{{\sc
-1}}}(p^{2})=\frac{1}{2}\int \frac{{\rm d}^{d}q}{(2\pi
)^{d}}\frac{1}{(q^{2}+m^{2})[(q+p)^{2}+m^{2}]}\ \cdot   \label{eq4}
\end{equation}
The partition function is then given by
\begin{equation}
Z=e^{-\frac{N}{2}S_{eff}(m^{2},g)}\int ({\cal {D}}\lambda )e^{-\frac{1}{2}
\int {\rm d}^{d}x\,{\rm d}^{d}y\bigl[\lambda (x)D(x-y)\lambda (y)\bigl]
+O(1/\sqrt{N})}\,,  \label{eq5}
\end{equation}
where, $D(x)$ is the $x$-space Fourier transform of (\ref{eq4}). The
effective theory above describes the ``propagation'' and ``interactions'' of
the composite field $\lambda (x)$. The critical theory, which is a non-trivial CFT
for $2<d<4$, is obtained for $
1/g\equiv 1/g_{\ast }=(2\pi)^{-d}\int{\rm d}^{d}p/p^{2}$ and $m\equiv M_{\ast }=0$.

When the system is put in a strip-like geometry having one finite dimension
of length $L$ and periodic boundary conditions, the momentum along the
finite dimension takes the discrete values $\omega _{n}=2\pi n/L$, $n=0,\pm
1,\pm 2,...$, and the relevant integrals become infinite sums. For example,
the gap equation and the inverse $\lambda (x)$ propagator become now
\begin{eqnarray}
\frac{1}{g} &=&\frac{1}{L}\,\!\!\sum\limits_{n=-\infty }^{\infty }\int \frac{
{\rm d}^{d-1}p}{(2\pi )^{d-1}}\frac{1}{p^{2}+\omega _{n}^{2}+m_{\text{{\sc L}
}}^{2}}\,,  \label{eq6} \\
\Pi _{\text{{\sc L}}}^{\text{{\sc -1}}}(p^{2},\omega _{n}^{2})
&=&\frac{1}{2L}\,\!\!\sum_{m=-\infty }^{\infty }\int \frac{{\rm d}
^{d-1}q}{(2\pi )^{d-1}}\frac{1}{(q^{2}+\omega _{m}^{2}+m_{\text{{\sc L}}
}^{2})[(q+p)^{2}+(\omega _{m}+\omega _{n})^{2}+m_{\text{{\sc L}}}^{2}]}\,\
\cdot  \label{eq7}
\end{eqnarray}
Since renormalisation of the bulk theory is sufficient for the
renormalisation of the theory in finite volume \cite{Zinn-Justin}, the
critical coupling in the latter case holds its bulk critical value
$1/g_{*}=(2\pi)^{-d}\int{\rm d}^{d}p/p^{2}$. The value of the critical
mass parameter $M_{\ast }$ however, may be different from zero. Indeed,
the (renormalised) gap equation for the finite system at criticality can
be found from (\ref{eq6}) to be
\begin{equation}
0=M_{\ast }^{d-2}\left[
\frac{\Gamma (\frac{1}{2}d-\frac{1}{2})\Gamma
(1-\frac{1}{2}d)}{4\sqrt{\pi }}+{\cal I}_{0}\right] ,  \label{eq10}
\end{equation}
where
\begin{equation}
{\cal I}_{n}=\int_{1}^{\infty }{\rm d}t\frac{(t^{2}-1)^{\frac{1}{2}\left(
d-3\right) +n}}{e^{LM_{\ast }t}-1}\quad \cdot
\end{equation}
For $2<d<4$, (\ref{eq10})
has a solution with $M_{\ast }\neq 0$. The dimensionless quantity
$M_{*}L$ is plotted for this case as a function of $d$ in Fig.1. For
$2<d\leq 3$, the massless phase is saturated, since (\ref{eq10})
diverges for $M_{\ast }$ zero, e.g. for $2<d\leq 3$ only the
$O(N)$-symmetric and critical theories exist. However, for $3<d<4$, (\ref
{eq10}) is also satisfied for $M_{\ast }=0$ and the model retains the
two-phase structure which has in the bulk. For $d=3$, which is a special
case as discussed below, the solution of (\ref{eq10}) takes the value $
M_{\ast }=(1/L)\ln \tau ^{2}$ \cite{Rosenstein, Sachdev}, where $\tau =(
\sqrt{5}+1)/2$ .

Before attempting the evaluation of the inverse $\lambda(x)$ propagator,
we briefly discuss the CFT approach to the finite size scaling of the
conformally invariant $O(N)$ vector model in $2<d<4$. This approach is
based on the OPE structure of the model, studied in a number of works
\cite {Ruhl,Tassos1,Tassos2}. The OPE of the basic field $\phi ^{\alpha
}(x)$ with itself takes the form
\begin{equation}
\phi ^{\alpha }(x)\phi
^{\beta }(0)=\frac{C_{\phi }}{x^{2\eta }}\delta ^{\alpha \beta
}+\frac{g_{\phi \phi {\cal O}}}{C_{\cal O}}\frac{1}{(x^{2})^{\eta
-\frac{1}{2} \eta _{o}}}{\cal O}(0)\delta ^{\alpha \beta }+\cdots ,
\label{eq11}
\end{equation}
where the dots stand for terms related to the
energy momentum tensor $T_{\mu\nu }(x)$, the $O(N)$ conserved current
$J_{\mu }^{\alpha \beta }(x)$ and other fields with less singular
coefficients as $x\rightarrow 0$. The field ${\cal O}(x)$ has dimension
$\eta _{o}=2+O(1/N)$, the coupling $ g_{\phi \phi {\cal O}}$ is
$O(1/\sqrt{N})$ and the dimension of $\phi ^{\alpha }(x)$ is $\eta
=d/2-1+O(1/N)$. $C_{\phi}$ and $C_{\cal O}$ are the normalisation
constants of the two-point functions of $\phi^{\a}(x)$ and
${\cal O}(x)$ respectively. Clearly, ${\cal O}(x)$ may readily be
identified with $\lambda(x)$ in (\ref{eq5}), we prefer however for the
sake of generality to make this identification at a later stage. The OPE of
${\cal O}(x)$ with itself takes the form \begin{equation} {\cal O}(x){\cal
O}(0)=\frac{C_{{\cal{O}}}}{x^{2\eta _{o}}}+\frac{g_{_{\cal O}}}{C_{\cal
O}}\frac{1}{ (x^{2})^{\frac{1}{2}\eta _{o}}}{\cal O}(0)+\frac{d\,\eta
_{o}\,C_{\cal O}}{ (d-1)S_{d}C_{T}}\frac{x_{\mu }x_{\nu }}{(x^{2})^{\eta
_{o}-\frac{1}{2}d+1}} T_{\mu \nu }(0)+\cdots ,  \label{eq12}
\end{equation}
where \cite{Tassos1} $C_{T}=N\,d/(d-1)S_{d}^{2}+O(1/N)$ is the normalisation
of the two-point function of $T_{\mu \nu }(x)$, with $S_{d}=2\pi
^{d/2}/\Gamma (d/2)$ the area of the $d-$dimensional unit sphere. The dots
stand for derivatives of ${\cal O}(x)$ and $T_{\mu \nu }(x)$, as well as for
other fields not relevant to the present work. The coefficient in front
of the leading $T_{\m\n}(x)$ contribution is exactly determined from
conformal invariance and the Ward identities satisfied by the three-point
function $\langle T_{\m\n}{\cal O}{\cal O}\rangle$ and we
also know that $g_{\cal O}C_{\phi}=2(d-3)g_{\phi \phi {\cal{O}}}C_{\cal
O}$ to leading $N$ \cite{Tassos1}.

Consider now the CFT having the OPEs above in the
strip-like geometry. Taking the expectation value of (\ref{eq12}) yields
the leading finite size corrections to the bulk-form $1/x^{2\eta _{o}}$ of
the two-point function of ${\cal O}(x)$. These arise from the fact that
the expectation values $\langle {\cal O}\rangle $ and $\langle T_{\mu \nu
}\rangle $ may in general be non-zero in a finite geometry. Given
the explicit form for ${\cal O}(x)$ in terms of the fundamental fields
$\phi^{\a}(x)$, one may explicitly calculate $\langle{\cal O}\rangle$ in a
$1/N$ expansion. Alternatively, the leading $N$ value for $\langle{\cal
O}\rangle$ can be also obtained by means of a consistency argument as we
show below. On the other hand, Cardy \cite {Cardy2} showed that for a
general conformal field theory the diagonal matrix elements \footnote{Th
off-diagonal matrix elements of $\langle T_{\mu \nu }\rangle $ vanish from
reflection symmetry.} of $\langle T_{\mu \nu }\rangle $ are related to the
finite size correction of its free energy density $f_{\infty}-f_{L}$. His
result reads
\begin{equation}
\langle T_{11}\rangle =-(d-1)\langle
T_{ii}\rangle =(d-1)\frac{2\zeta (d)}{
S_{d}L^{d}}\tilde{c}=(d-1)(f_{\infty}-f_{L})\hspace{1cm}\mbox{(no summation
on}\,\,i\,)\,,  \label{eq13}
\end{equation}
where, $\tilde{c}$ is a universal number which has
been considered to be a candidate for a possible generalisation of
Zamolodchikov's $C$-function for $d>2$ \cite{Fradkin}.

Upon transforming (\ref{eq12}) to momentum space for $\eta _{o}=2$ and
using (\ref{eq13}), the leading finite size corrections for the ${\cal
O}(x)$ propagator  can be written as
\begin{eqnarray}
{\cal
O}_{\text{{\sc L}}}(p^{2},\omega _{n}^{2}) &=&\frac{ C_{\cal O}\,\pi
^{\frac{1}{2}d}\,2^{d-4}\,\Gamma ({\textstyle{\frac{1}{2}d-2}})}{
(p^{2}+\omega _{n}^{2})^{\frac{1}{2}d-2}}\Biggl(1+4\left( {\textstyle{
\frac{d}{2}-2}}\right)\frac{g_{\cal O}}{C_{\cal O}^{2}}
\frac{1}{p^{2}+\omega _{n}^{2}}\langle {\cal O} \rangle   \nonumber \\
&&\hspace{2cm}+\frac{2^{d+1}\Gamma (d-1)S_{d}}{\pi ^{\frac{1}{2}d}\,\Gamma
(2-\frac{1}{2}d)}\,\frac{\zeta (d)\tilde{c}}{L^{d}\,N}\,\frac{C_{2}^{
\scriptstyle{\frac{1}{2}d-1}}(y)}{(p^{2}+\omega _{n}^{2})^{\frac{1}{2}d}}
+....\Biggl),  \label{eq14}
\end{eqnarray}
where $C_{\nu }^{\lambda }(y)$ are the Gegenbauer polynomials and $y=\omega
_{n}/\sqrt{p^{2}+\omega _{n}^{2}}$ . For $2<d<4$, the terms shown in (\ref
{eq14}) are the most singular ones in the large momentum expansion of
${\cal O}_{L}(p^{2},\omega _{n}^{2})$.

Similarly, taking the expectation value of (\ref{eq11}) yields the
leading  finite size corrections to the two-point function of
$\phi^{\a}(x)$. Then, by transforming (\ref{eq11}) to momentum space
we obtain for the propagator $P_{\text{{\sc L}}}(p^{2},\omega_{n}^{2})$ of
$\phi^{\a}(x)$
\begin{equation}
P_{\text{{\sc L}}}(p^{2},\omega
_{n}^{2})=C_{\phi }\frac{4\pi ^{\frac{1}{2}d}}{\Gamma
(\frac{1}{2}d-1)}\frac{1}{p^{2}+\omega _{n}^{2}}\left( 1+4\left(
{\textstyle{\frac{d}{2}-2}}\right) \frac{g_{\phi \phi {\cal
O}}}{C_{\phi}C_{\cal O}} \frac{1}{p^{2}+\omega _{n}^{2}}\langle {\cal
O}\rangle +\cdots \right) . \label{eq15}
\end{equation}

If ${\cal O}(x)$ is to be identified now with the composite field
$\lambda(x)$ in (\ref{eq5}), formulae (\ref{eq14}) and (\ref{eq15}) must
be consistent with what is expected by explicitly calculating the
propagators of $\lambda(x)$ and $\phi^{\a}(x)$, in the
context of the standard $1/N$ expansion at the critical point of the
$O(N)$ vector model. Here, we will only be concerned with leading $N$
calculations and we choose to work in momentum space as it is customary for
finite size scaling studies \cite{Sachdev,Gatto}.

We start our consistency analysis from $P_{\text{{\sc
L}}}(p^{2},\omega_{n}^{2})$ which, to leading $N$, is easily found from
(\ref{eq1}) to be
\begin{equation}
P_{\text{{\sc L}}}(p^{2},\omega
_{n}^{2})=\frac{1}{ p^{2}+\omega _{n}^{2}+M_{\ast
}^{2}}=\frac{1}{p^{2}+\omega _{n}^{2}}\left( 1- \frac{M_{\ast
}^{2}}{p^{2}+\omega _{n}^{2}}+...\right) \ \cdot   \label{eq16}
\end{equation}
Note that, to this order there is no contribution from $T_{\m\n}(x)$ to
the rhs of (\ref{eq16}). Consistency of (\ref{eq16}) with (\ref{eq11})
requires that $C_{\phi }=\Gamma (d/2-1)/4\pi ^{d/2}$ and also
\begin{equation}
\langle {\cal O}\rangle =-\frac{M_{\ast
}^{2}C_{\phi}C_{\cal O}}{4g_{\phi \phi{\cal O}}(\frac{1}{2}
d-2)}=-\frac{M_{\ast }^{2}(d-3)C_{\cal
O}^{2}}{2g_{{\cal{O}}}(\frac{1}{2}d-2)}\ \cdot \label{eq17} \end{equation}

Next, we turn to the propagator $\Pi_{\text{{\sc L}}}^{\text{{\sc
-1}}}(p^{2},\omega_{n}^{2})\,(\equiv {\cal{O}}^{-1}_{L}(p^{2}
,\omega_{n}^{2}))$ of $\lambda(x)$, in (\ref{eq7}). Using standard algebra
we obtain \begin{eqnarray} {\cal O}_{\text{{\sc L}}}^{\text{{\sc
-1}}}(p^{2},\omega _{n}^{2}) &=&\frac{2^{2-d}\Gamma
({\textstyle{2-\frac{1}{2}d}})\pi ^{\frac{1}{2}d}}{ (2\pi
)^{d}}(p^{2}+\omega _{n}^{2})^{\frac{1}{2}d-2}{}_{2}F_{1}\left( {
\textstyle{2-\frac{1}{2}d,\frac{1}{2};\frac{3}{2};\frac{p^{2}+\omega _{n}^{2}
}{p^{2}+\omega _{n}^{2}+4M_{\ast }^{2}}}}\right)   \nonumber \\
&&\hspace{-2.5cm}+\int \frac{{\rm d}^{d-1}q}{\left( 2\pi \right) ^{d-1}}
\frac{1}{\sqrt{q^{2}+M_{\ast }^{2}\,}\left[ e^{L\sqrt{q^{2}+M_{\ast }^{2}\,}
}-1\right] }\frac{p^{2}+\omega _{n}^{2}+2q\!\cdot \!p}{\left( p^{2}+\omega
_{n}^{2}+2q\!\cdot \!p\right) ^{2}+4\omega _{n}^{2}\left( q^{2}+M_{\ast
}^{2}\right) }\hspace{-3cm}\ \cdot   \label{plmin}
\end{eqnarray}
Introducing the quantities
\begin{equation}
A_{d}=\frac{2^{2-d}\pi ^{\frac{1}{2}d+\frac{1}{2}}\Gamma (2-\frac{1}{2}
d)\Gamma (\frac{1}{2}d-1)}{(2\pi )^{d}\Gamma (\frac{1}{2}d-\frac{1}{2})}
\,\,\,\,\mbox{and}\,\,\,\,v^{2}=\frac{4M_{\ast }^{2}}{p^{2}+\omega _{n}^{2}}
\ ,
\end{equation}
the first term on the rhs of (\ref{plmin}) can be written in a form more
suitable for  large momentum expansion as,
\begin{equation}
A_{d}(p^{2}+\omega _{n}^{2})^{\frac{1}{2}d-2}\left[ \left( 1+v^{2}\right) ^{
\frac{1}{2}d-\frac{3}{2}}+\frac{\Gamma (\frac{1}{2}d-\frac{1}{2})}{\sqrt{\pi
}\Gamma (d)}\frac{(v^{2})^{\frac{d}{2}-1}}{1+v^{2}}{}_{2}F_{1}\left(
\textstyle{{\frac{1}{2}d-\frac{1}{2},1;\frac{1}{2}d;\frac{v^{2}}{1+v^{2}}}}
\right) \right] \,\ \cdot   \label{hyp1}
\end{equation}
The second term on the rhs of (\ref{plmin}) can be written as
\begin{eqnarray}
&&\frac{S_{d-1}}{(2\pi )^{d-1}}\!\!\int_{0}^{\infty }\!\frac{q^{d-2}{\rm d}q
}{(q^{2}+M_{\ast }^{2})^{\frac{1}{2}d}\left[ e^{L\sqrt{q^{2}+M_{\ast }^{2}\,}
}-1\right] }\mbox{Re}\left[ I(p^{2},\omega _{n}^{2},q)\right] ,
\label{angp} \\
&&I(p^{2},\omega _{n}^{2},q)=\sum_{k=0}^{\infty }\frac{
(-2i)^{k}\,r^{k}}{(p^{2}+\omega _{n}^{2})^{k+1}}\!\int_{0}^{\pi }\!\!\sin ^{
\scriptstyle{d-3}}\!\phi \left( \cos t-i{\textstyle}\sin t\,\cos \phi
\right) ^{k}\!\!\!\ {\rm d}\phi ,
\end{eqnarray}
where, we have defined
\begin{equation}
\omega _{n}\sqrt{q^{2}+M_{\ast }^{2}}=r\,\cos t,\quad q\,p=r\,\sin t,\quad r=
\sqrt{\left( p^{2}+\omega _{n}^{2}\right) q^{2}+\omega _{n}^{2}M_{\ast }^{2}}
\ \cdot
\end{equation}
The angular integral in (\ref{angp}) can be evaluated \cite{Gradshteyn} in
terms of Gegenbauer polynomials and, taking the real part, we obtain
\begin{equation}
{\rm{Re}}\left[ I(p^{2},\omega _{n}^{2},q)\right]
=\sum_{k=0}^{\infty }\frac{(-4)^{k}\Gamma ^{2}(\frac{1}{2}d-1)2^{d-3}(2k)!}{
\Gamma (d-2+2k)}\frac{r^{2k}}{(p^{2}+\omega _{n}^{2})^{2k+1}}C_{2k}^{\frac{1
}{2}d-1}\left( \cos t\right) \ \cdot   \label{eq19a}
\end{equation}
It is evident that (\ref{eq19a}) is a large momentum expansion. The leading
terms of ${\cal O}_{\text{{\sc L}}}^{\text{{\sc
-1}}}(p^{2},\omega_{n}^{2})$ can now be evaluated after substituting
(\ref{eq19a} ),(\ref{hyp1}) in (\ref{plmin}). Apparently, the
hypergeometric function in (\ref{hyp1}) contributes terms which do not
seem to be present in (\ref{eq14}). The second term of (\ref{hyp1}) has to
be interpreted as a dimension $d-2$ scalar field contribution. This field
corresponds to the ``shadow field''\footnote{ The ``shadow field'' of a
scalar field $A(x)$ with dimension $l$ is a scalar field with dimension
$d-l$. For the ``shadow symmetry'' of the conformal group in $d>2$ see
\cite{Parisi} and references therein.} of ${\cal O}(x)$. Note that the
dynamics of the $O(N)$ vector model requires that ``shadow singularities''
cancel out from four-point functions \cite{Tassos1}. In order to clarify
the role of the ``shadow singularities'' inside the inverse two-point
function (\ref{plmin}), we must reexpress (\ref {eq19a}) in terms of
$C_{2k}^{d/2-1}(y)$. Given that $C_{\nu }^{\lambda }(y)$ for integer $\nu
$ are orthogonal polynomials of order $\nu $, an expansion of the form
\begin{equation}
(q^{2}+y^{2}M_{\ast }^{2})^{k}\,C_{2k}^{\frac{1}{2}d-1}\left( \frac{y\left(
q^{2}+M_{\ast }^{2}\right) }{\sqrt{q^{2}+y^{2}M_{\ast }^{2}}}\right)
=\sum_{l=0}^{2k}B_{l}\,(q^{2},M_{\ast }^{2})\,C_{l}^{\frac{1}{2}d-1}(y)
\label{eq21a}
\end{equation}
exists. The scalar contribution $B_{0}(q^{2},M_{*}^{2})\equiv
B_{0}(M_{*}^{2})$ can be easily evaluated and reads \begin{equation}
B_{0}(M_{*}^{2})= \frac{\Gamma (\frac{d}{2})\,\Gamma (d-2+2k)\,\Gamma
(k+\frac{1}{2})}{ \sqrt{\pi }\,\Gamma (d-2)\,\Gamma (2k+1)\,\Gamma
(k+\frac{d}{2})}M_{\ast }^{2k}\ \cdot   \label{eq21aa} \end{equation}
Substituting $B_{0}$ in (\ref{plmin}) we obtain, after some algebra,
\begin{equation}
-\frac{4}{\Gamma (\frac{1}{2}d)\Gamma (1-\frac{1}{2}d)}\frac{(v^{2})^{\frac{1
}{2}d-1}}{1+v^{2}}{}_{2}F_{1}\left( {\textstyle{1,\frac{1}{2}d-\frac{1}{2};
\frac{1}{2}d;\frac{v^{2}}{1+v^{2}}}}\right) \,{\cal I}_{0}\ ,
\label{eq21aaa}
\end{equation}
which cancels exactly the second term in (\ref{hyp1}) by virtue of the gap
equation (\ref{eq10}).

Thus, the gap equation (\ref{eq10}), being a necessary and sufficient
condition for this cancellation to occur, suggests an interesting
alternative approach for studying this model. Namely, we could have started
with (\ref{eq7}) as the definition of the inverse two-point function of the
field $\lambda (x)$. Requiring then that this is an inverse two-point
function of a CFT obtainable from an OPE, and, postulating that no ``shadow
singularities'' appear we are led, by means of the same calculations, to the
gap equation (\ref{eq10}). Consequently, (\ref{eq10}), which is the basic
dynamical equation of the $O(N)$ model, could be viewed as a precondition
for the ``shadow singularities'' cancellation. Such an algebraic approach to
CFT in $d>2$ was initiated in \cite{Tassos2} and it is now demonstrated to
work for finite geometries as well.

Eventually, the first few terms in the large momentum expansion of
(\ref{plmin}) are \begin{eqnarray} {\cal O}_{\text{{\sc L}}}^{\text{{\sc
-1}}}(p^{2},\omega_{n}^{2}) &=&A_{d}(p^{2}+\omega
_{n}^{2})^{\frac{1}{2}d-2}\left[ 1+{\textstyle{2(d-3)}} \frac{M_{\ast
}^{2}}{p^{2}+\omega _{n}^{2}}\right.   \nonumber \\ &&\left.
+\frac{2^{2d}\Gamma (\frac{1}{2}d-\frac{1}{2})}{\sqrt{\pi }\Gamma
(d)\Gamma (2-\frac{1}{2}d)}\frac{C_{2}^{\frac{1}{2}d-1}(y)}{(p^{2}+\omega
_{n}^{2})^{\frac{1}{2}d}}M_{\ast }^{d}\,\left( {{\frac{1-d}{d}}}{\cal {I}}
_{0}-{\cal {I}}_{1}\right) +\cdots \right]\,.   \label{eq22}
\end{eqnarray}
Now, (\ref{eq22}) must be consistent with (\ref{eq14}). It is then easy
to see that the $O(M_{*}^{2})$ terms on the rhs of (\ref{eq14}) and
(\ref{eq22}) coincide by virtue of (\ref{eq17}). This is a non-trivial
consistency check, since it requires the non-trivial relation between the
couplings $g_{\cal O}$ and $g_{\phi\phi{\cal O}}$. Next, consistency  of
the angular terms proportional to $C_{2}^{d/2-1}(y)$ in (\ref{eq14}) and
(\ref{eq22}) requires that \begin{equation} \Gamma (d)\zeta
(d)\frac{\tilde{c}}{L^{d}N}=M_{\ast }^{d}\,\left( {{\frac{d-1 }{d}}}{\cal
{I}}_{0}+{\cal {I}}_{1}\right) \,.  \label{eq25} \end{equation} In order
to prove that, we calculate the free energy density of the theory
(\ref{eq1}) to leading order in $1/N$
\begin{equation}
\hspace{-0.5cm}f_{\infty }-f_{\text{{\sc L}}}\equiv \frac{2\zeta (d)}{
S_{d}L^{d}}\tilde{c}=N\,M_{\ast }^{d}\frac{S_{d-1}}{(2\pi )^{d-1}}\left( {{
\frac{1}{d}}}{\cal {I}}_{0}-{{\frac{1}{LM_{\ast }}}}\int_{1}^{\infty
}\!x\,(x^{2}-1)^{\frac{1}{2}d-\frac{3}{2}}\,\ln (1-e^{-LM_{\ast }x}){\rm d}
x\right) \cdot   \label{eq26}
\end{equation}
Now, for $2<d<4$, a partial integration yields
\begin{equation}
{\cal I}_{1}=-{\textstyle{\frac{d-1}{LM_{\ast }}}}\int_{1}^{\infty
}x\,(x^{2}-1)^{\frac{1}{2}d-\frac{3}{2}}\,\ln (1-e^{-LM_{\ast }x}){\rm d}x\
\cdot   \label{eq27}
\end{equation}
Then, using the identity $S_{d}\,S_{d-1}=2\,(2\pi )^{d-1}/\Gamma (d-1)$ one
can see that (\ref{eq25}) is satisfied by means of (\ref{eq26}) and (\ref
{eq27}). To the best of our
knowledge, this is the first time that the validity of Cardy's result
\cite{Cardy2} is explicitly demonstrated for CFTs in $d>2$. For
completeness, a plot of $\tilde{c}/N$ as a function of $d$ for $2<d<4$ is
depicted in Fig.2. The evaluation of $\tilde{c}$ to $O(1/N)$, and its
relation to possible generalisations of Zamolodchikov's $C$-function for
$d>2$, are given in \cite{Tassos3}.

The case $d=3$ is special. From (\ref{eq14}) or (\ref{plmin}), we see that
there in no self-contribution from ${\cal O}(x)$  to its
two-point function, which is related to the fact that the bulk theory
respects the reflection symmetry property ${\cal{O}}(x)\rightarrow
-{\cal{O}}(x)$ to leading order in $1/N$. 
Also, for $d=3$, $M_{\ast }=(1/L)\ln \tau ^{2}$ and then it can be
shown \cite{Tassos3} that the integrals involved in (\ref{eq26}) are related
to polylogarithms at the special point $2-\tau $, leading to the
surprisingly simple result $\tilde{c}/N=4/5$ \cite{Sachdev}.

For $3<d<4$, the gap equation (\ref{eq10}) has also the solution $
M_{\ast }=0$. In this case, the corresponding expression for ${\cal
O}_{\text{{\sc L}}}^{\text{{\sc -1}}}(p^{2},\omega_{n}^{2})$ is given by
\begin{eqnarray}
&&\hspace{-1cm}{\cal O}_{\text{{\sc L}}}^{ \text{{\sc
-1}}}(p^{2},\omega_{n}^{2})=  \nonumber \\
&&\hspace{1.1cm}=A_{d}(p^{2}+\omega _{n}^{2})^{\frac{1}{2}d-2}\left( 1+\frac{
2^{2d-3}\Gamma (\frac{1}{2}d-\frac{1}{2})}{\sqrt{\pi }\Gamma (2-\frac{1}{2}d)
}\!\!\sum_{k=0}^{\infty }\frac{(-4)^{k}(2k)!\,\zeta (d-2+2k)}{
[L^{2}(p^{2}+\omega _{n}^{2})]^{\frac{1}{2}d-1+k}}C_{2k}^{\frac{1}{2}
d-1}(y)\right)  \nonumber \\
&&\hspace{1.1cm}=\,A_{d}\,(p^{2}+\omega _{n}^{2})^{\frac{1}{2}d-2}\left( 1+
\frac{2^{2d-3}\,\Gamma (\frac{1}{2}d-\frac{1}{2})}{\sqrt{\pi }\,\Gamma (2-
\frac{1}{2}d)}\frac{\zeta (d-2)}{[L^{2}(p^{2}+\omega _{n}^{2})]^{\frac{1}{2}
d-1}}\right.  \nonumber \\
&&\hspace{2cm}\left. -\frac{2^{2d}\Gamma (\frac{1}{2}d-\frac{1}{2})}{\sqrt{
\pi }\Gamma (2-\frac{1}{2}d)}\frac{\zeta (d)}{[L^{2}(p^{2}+\omega
_{n}^{2})]^{\frac{1}{2}d}}C_{2}^{\frac{1}{2}d-1}(y)+\cdots \right) \ \cdot
\label{eq28}
\end{eqnarray}
Since now the propagator of $\phi ^{\alpha }(x)$ is
simply $1/(p^{2}+\omega _{n}^{2})$, consistency with (\ref{eq13})
requires $\langle {\cal O}\rangle =0$ to leading order in $1/N$ which
means that we do not expect to get any contributions from $ {\cal O}(x)$
in $\Pi _{\text{{\sc L}}}^{\text{{\sc -1}}}(p^{2},\omega _{n}^{2};M_{\ast
}^{2})$. Indeed, a term $\propto 1/(p^{2}+\omega _{n}^{2})^{d/2-1}$ is
absent from the rhs of (\ref{eq28}). Instead, we find a term $\propto
1/(p^{2}+\omega _{n}^{2})$ which has the correct dimensions to be
interpreted as the leading contribution of the ``shadow field'' of ${\cal
 O}(x)$. This term is not cancelled out and this is a characteristic
feature of free field theories as discussed in \cite{Tassos1}.

Next, we require consistency of the coefficients of
$C_{2}^{d/2-1}{\textstyle{(y)}}$ in (\ref{eq28}) and (\ref{eq14}) which
yields, after some algebra, $\tilde{c}=N\ $. This is exactly the result
one obtains by calculating $f_{\infty }-f_{\text{{\sc L}}}$ to leading $N$
at the critical point (with $ M_{\ast }=0)$, i.e. the result for free
massless scalar fields. Note that (\ref{eq28}) looks
like a free field theory decomposition for the two-point function of
${\cal O} (x)$, ${\cal O} (x)$ which has in the place of ${\cal O} (x)$ the
dimension $d-2$ composite scalar field $:\phi ^{2}(x):$.

Our results above provide evidence that two-point functions of CFTs in
$d>2$ and in finite geometry are determined by bulk conformally invariant
OPEs. In fact, the dynamical equations for the finite size theory are just
the conditions for the cancellation of \ ``shadow'' singularities as shown
in (\ref{hyp1}) and (\ref{eq21aaa}). Furthermore, we have explicitly shown
that the leading angular correction to the scalar two-point function is
proportional to $ \tilde{c}/C_{T}$. It would be interesting to look for
non trivial cancellations inside the two point function of $\phi ^{\alpha
}(x)$ where the calculations will be related to those in \cite{Sachdev}
and \cite{Rossi} . Extension of the OPE approach to fermionic, $CP^{N-1}$
and supersymmetric CFTs in $2<d<4$ would further clarify the connection
between ``shadow singularity'' cancellations and dynamics. Another
issue might be whether the conformal theory with $M_{*}=0$ is a free
field theory, which could be clarified by next-to-leading order in
$1/N$ calculations.

{\it This work was supported in part by PENED/95 K.A. 1795\ research \ grant.}

\newpage

%\begin{center}
%{\Large {\bf {Figure Captions} }}
%\end{center}

%{\normalsize {\bf Fig.1: \ }The mass gap $M_{\ast }$ as a function of the
%spacetime dimensionality $d.$ }

%{\normalsize {\bf Fig.2: \ }Plot of $\tilde{c}/N$ as a function of the
%spacetime dimensionality $d.$ }

\begin{figure}[h]
\epsfxsize=6cm
\centerline{\epsffile{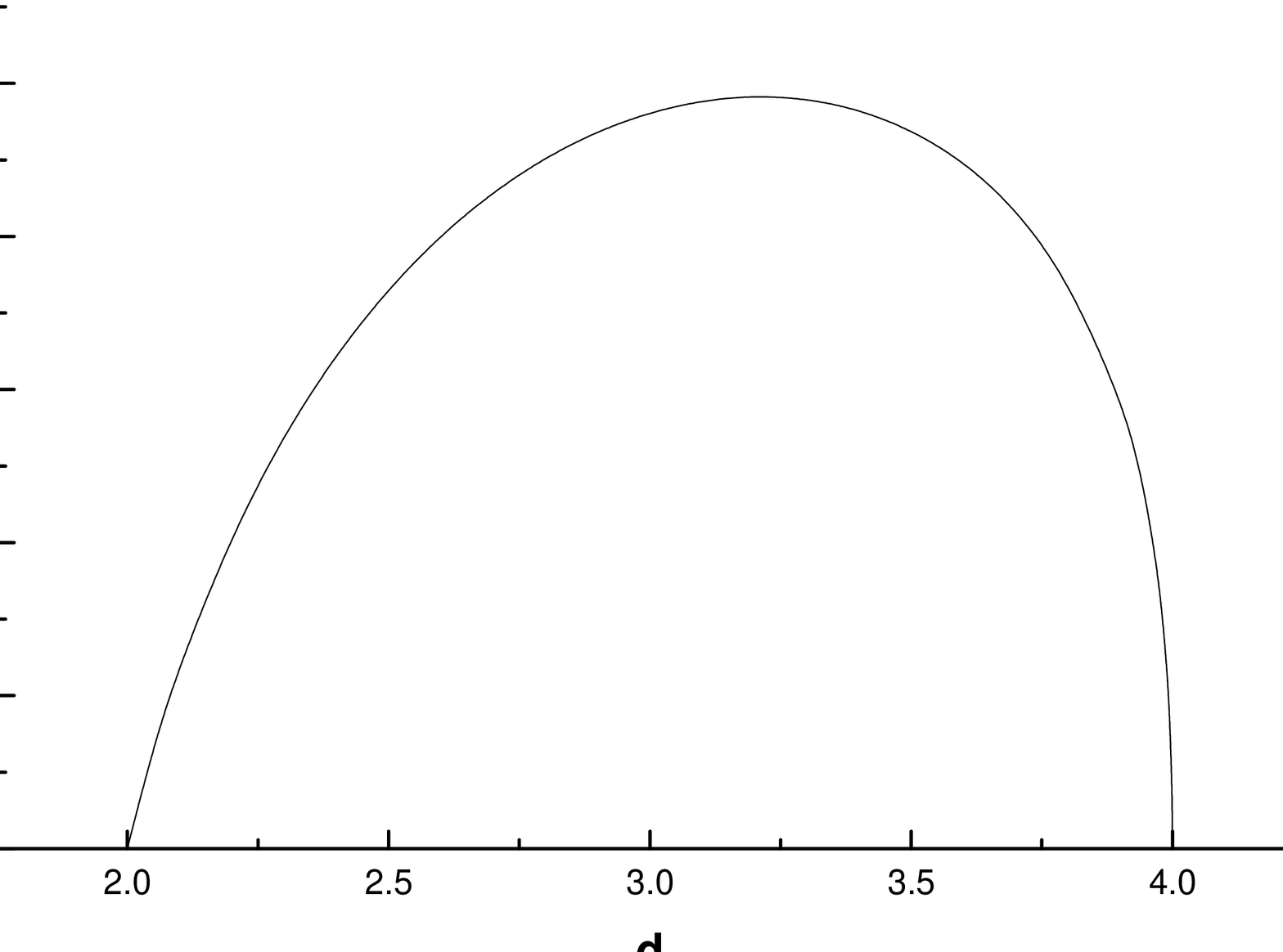}}
\caption{The dimensionless quantity $M_{\ast }L$ as a function of the
spacetime dimensionality $d.$ }
\end{figure}

\hspace{6cm}

\begin{figure}[h]
\epsfxsize=6cm
\centerline{\epsffile{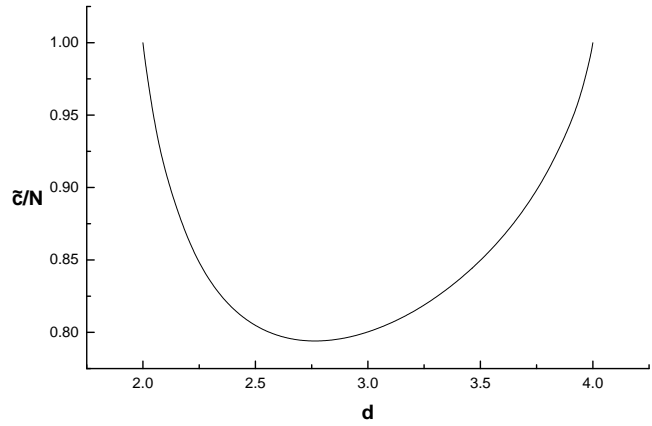}}
\caption{Plot of $\tilde{c}/N$ as a function of the
spacetime dimensionality $d.$}
\end{figure}

\end{document}